# Optimising the Trade-Off Between Type I and Type II Errors: A Review and Extensions.

Andrew P Grieve[1,2]

**Abstract**

In clinical studies upon which decisions are based there are two types of errors that can be made: a type I error arises when the decision is taken to declare a positive outcome when the truth is in fact negative, and a type II error arises when the decision is taken to declare a negative outcome when the truth is in fact positive. Commonly the primary analysis of such a study entails a two-sided hypothesis test with a type I error rate of 5% and the study is designed to have a sufficiently low type II error rate, for example 10% or 20%. These values are arbitrary and often do not reflect the clinical, or regulatory, context of the study and ignore both the relative costs of making either type of error and the sponsor's prior belief that the drug is superior to either placebo, or a standard of care if relevant. This simplistic approach has recently been challenged by numerous authors both from a frequentist and Bayesian perspective since when resources are constrained there will be a need to consider a trade-off between type I and type II errors. In this paper we review proposals to utilise the trade-off by formally acknowledging the costs to optimise the choice of error rates for simple, point null and alternative hypotheses and extend the results to composite, or interval hypotheses, showing links to the Probability of Success of a clinical study.

**1. Introduction.**   -

The recent debate on the appropriate level of evidence required for claiming new findings in many scientific endeavours has been wide-ranging and several proposals have been made to modify standard practice. For example, Benjamin et al (2018) have proposed a blanket adoption of a more stringent criterion for judging the success of an experiment, namely, to use a type I error rate of 0.005 instead of the almost universally used value of 0.05. Whilst recognising the abuse of the term statistical significance which should lead to discontinuation of its use, Lakens et al (2018) propose that each individual scientist, or research team, should determine the appropriate level of type I error against which evidence should be judged for the specific context of their research. In this paper we lean towards the "variable proposal" of Lakens et al (2018) rather than the "blanket proposal" of Benjamin et al (2018), despite it not being universally welcomed (Trafimow, 2018). The principal reason for our choice is that it acknowledges context. Furthermore, we believe that the approach to in this paper provides a solid scientific rationale for the choice of type I and type II error rates, in contrast to a "blanket proposal" which leads to essentially arbitrary values. Several research teams have proposed to study the trade-off between the type I and type II error rates, in contrast to what is generally done which is to fix the type I error rate and then design the study to ensure that the power of the study to detect meaningful effect sizes is sufficiently large.

Mudge et al (2012) investigated the potential trade-off between the type I and type II error for a fixed sample size in the context of frequentist significance testing. They argue that the type I error should be selected to minimise a weighted sum of the error rates and that their approach provides a principled methodology for choosing the type I error, which provides one method to implement a "variable" proposal. In their proposal Mudge et al concentrate on the use of weights which are related to the costs, or the relative costs of making type I and type II errors. Suppose that the cost of a type I error is $\omega_1$ and that of a type II error is $\omega_2$ then the weighted sum of error rates is,

---

1  *Pharmaceutical Sciences Research Division, King's College London, Franklin-Wilkins Building, 150 Stamford Street, London, SE1 9NH, United Kingdom; email: k2371396@kcl.ac.uk*
2  *This work was carried out while the author was a member of Statistical Innovation, UCB Pharma, 208 Bath Road, Slough, Berkshire SL2 3WE, UK.*

$$\frac{\omega_1 \alpha + \omega_2 \beta}{\omega_1 + \omega_2} = \frac{\omega \alpha + \beta}{\omega + 1} \tag{1}$$

where $\omega = \omega_1/\omega_2$ so that in effect they are proposing to minimise the overall expected cost of a wrong decision. They also consider weights which reflect the prior probabilities of the null and alternative hypotheses. In that context they argue that absent reliable information to form the basis of the prior probabilities it is rationale to use the principle of insufficient reason standpoint and to take $p_1 = p_2 = 0.5$, giving a weighted average of $(\alpha + \beta)/2$.

Mallinckrodt et al (2012) who consider a similar problem, assume that a clinical trial is being designed to provide evidence of the efficacy of a drug in which as before the type I error rate is $\alpha$ and the type II error rate is $\beta$. Additionally, they assume that the cost of a type I error is $C_\alpha$, and that of a type II error is $C_\beta$. If the probability that the drug is effective is $P(E)$, then the expected cost of a false positive is,

$$P(pos|neg)P(neg)cost(False\ Positive) = \alpha(1 - P(E))C_\alpha.$$

Similarly, the expected cost of a false negative is $\beta P(E) C_\beta$ so that the total expected cost is,

$$\Psi = \alpha(1 - P(E))C_\alpha + \beta P(E) C_\beta = \omega_1 \alpha + \omega_2 \beta \tag{2}$$

where $\omega_1 = C_\alpha(1 - P(E)), \omega_2 = C_\beta P(E)$

Isakov et al (2019) propose a more complex cost model for clinical trials. They distinguish between in-trial and post-trial costs, the former depending on the number of recruited patients, the latter depending on the results of the trial only. They assume that the post-trial costs associated with making a type I or a type II error are $c_1$ and $c_2$ per individual in the target populations, numbering $N$. In contrast, the in-trial costs are principally related either to a patient being exposed to a treatment with poor efficacy, or a delay in treating the entire target population with an efficacious treatment. For example, if the test drug has insufficient efficacy, patients in the treatment arm experience a total in-trial cost of $n \times c_1$, patients in the control arm experience no additional cost. In contrast, for an efficacious drug an inappropriately larger sample size would result in a delay in treating the target population with the efficacious treatment. The delay impacts all patients, whether in the trial or not, and Isakov et al (2019) model it as a fixed fraction, $\gamma$, of the total type II error cost $N \times c_2$, that is $n \times \gamma \times N \times c_2$. Finally, they assume that $p_0, p_1 > 0$ are the prior probabilities associated with the null and alternative hypotheses respectively and $p_0 + p_1 = 1$.

With these of assumptions the expected cost of the trial is

$$Cost(\alpha, n) = p_0 c_1 N\alpha + p_1 c_2 N\beta + p_0 c_1 n + p_1 c_2 n\gamma N.$$

Isakov et al (2019) simplify the model by assuming an uninformative prior, $p_0 = p_1 = 0.5$ implying that the total cost is proportional to

$$Cost(\alpha, n) = c_1 N\alpha + c_2 N\beta + c_1 n + c_2 n\gamma N$$

and minimise this expression over $\alpha$ and $n$, subject to a bound on the power of the study $1 - \beta$. Their approach is to find the optimum by fixing the sample size n and determining the optimal $\alpha$ and then optimising over $n$. For fixed $n$, the expected cost as a function of $\alpha$ is,

$$\Psi = c_1 \alpha + c_2 \beta \tag{3}$$

All three of these independent developments lead to the minimisation of a weighted sum of the type I and type II error rates given by (1), (2) and (3). The weights in each case are slightly different, involving either the costs of the individual errors, or the probability of the hypotheses, or both. The basic idea of this paper is to choose that pair of $\alpha$ and $\beta$ which minimise the total cost.

Savage and Lindley (Savage, 1961; Savage, 1962; Lindley, 1972) were the first to propose minimising a weighted average of the type I and type II error rates. Their work was aimed at deciding which amongst all possible pairs of $\alpha$ and $\beta$ to choose. They showed that amongst indifference curves the only possible structure was a series of parallel straight lines in which the slope is the negative ratio of the priors of the null and alternative hypotheses, and that this provides the critical likelihood ratio for any trial comparing a simple null hypothesis with a simple alternative hypothesis (see also Grieve, 2015) which Savage calls a "simple dichotomy". Arrow (1960) proposed an "equal-probability" test in which $\alpha = \beta$, corresponding to equal weights. Cornfield (1966), influenced by his correspondence with Savage (Greenhouse, 2012), proposed to minimise a linear combination of $\alpha$ and $\beta$ in which the slope parameter represented the cost, of $\alpha$ relative to $\beta$ and noted that the accept/reject region must consist of all likelihood ratio values that are less/greater than the given slope. He shows that the same principle applies to sequential trials. More modern authors have reported the same results. For example, DeGroot (1975) has argued that minimising a weighted sum of $\alpha$ and $\beta$ "is more reasonable" than to fix $\alpha$ and then to minimise $\beta$. Bernardo and Smith (1994) take a similar stance, arguing that minimising a linear combination of $\alpha$ and $\beta$ provides the only coherent approach as it corresponds to minimising the expected loss, unlike other procedures.

In Section 2 we review the work of Grieve (2015), who developed an analytic solution to the minimisation of the weighted sum of error rates for the comparison of simple point null and alternative hypotheses from a frequentist perspective for normally distributed data with known variance and Walley and Grieve (2021) the same from a Bayesian perspective. In Section 3 we generalise this approach to composite, interval hypotheses and show how there are links to the concepts of assurance and Probability of Success (*PoS*). In section 4 we provide an example of the determination of the relative cost associated with the type I and type II errors. The main results are given as a series of theorems and lemmas, whose proofs are provided in the appendix.

## 2. Balancing Type I and Type II Errors When the Hypotheses are Simple.
--
For illustrative purposes, we will assume throughout that we are interested in designing a two-arm clinical study in which the primary variable is normally distributed with known variance, $\sigma^2$. Furthermore, we will assume that the effect size we will be interested in detecting is $\delta_0$, the one-sided significance level (probability of a type I error or false positive rate) is $\alpha$ and the probability of a type II error (or false negative rate) is $\beta$.

### 2.1 Frequentist Analysis.

Under these simple assumptions, the sample size in each arm, $n_1$, to test the simple null hypothesis $H_o: \delta = 0$ against the simple alternative hypothesis $H_A: \delta = \delta_0$ with significance level, $\alpha$, and power, $1 - \beta$, is given by the formula,

$$n_1 = \frac{2\sigma^2(Z_{1-\alpha} + Z_{1-\beta})^2}{\delta_0^2} \qquad (4)$$

Although not appropriate in all circumstances, the central limit theorem allows us to apply this formula to many cases, at least after a suitable transformation. For our purposes we assume its use is appropriate.

**Example 1: Designing a Two-Arm Clinical Trial in Restless Leg Syndrome**

We use the planning of a clinical trial introduced by Muirhead and Soaita (2013) throughout this paper to illustrate the concepts. In their paper, Muirhead and Soaita considered the design of a trial to compare a new drug with a placebo in the treatment of Restless Leg Syndrome (RLS). The primary endpoint of the trial was the change from baseline of the scores on the International Restless Leg Scale (IRLS). Based on data published by Allen et al (2010), they assumed a standard deviation of 8 IRLS units and that a clinically meaningful treatment difference was $\delta = 4$ IRLS units. Designing the trial to achieve a type II error rate of 20% with a one-sided type I error rate of 2.5% it requires from (4)

$$n_1 = \frac{2 \times 64 \times (1.960 + 0.841)^2}{16} = 62.8$$

giving a sample size of 63 patients per arm which Muirhead and Soaita increased to 64.

Given values for $n_1, \alpha$ and $\delta_0$ the relationship between the probability of a type II error to test $H_o$ against $H_A$ and the probability of a type I error can be obtained by re-writing (4) to give,

$$\beta = 1 - \Phi(\theta + Z_\alpha) \tag{5}$$

,

$$\theta = \sqrt{n_1/2}\, \delta_0/\sigma \tag{6}$$

is the non-centrality parameter. For a given $\omega$, either the "relative prior probabilities of the null and alternate hypotheses being true" (Grieve, 2015) or the relative costs of the individual errors, the weighted sum of the probabilities of type I and type II errors is given by,

$$\Psi = \frac{\omega\alpha + 1 - \Phi(\theta + Z_\alpha)}{\omega + 1} \tag{7}$$

which, for fixed $\omega$, is a function of $\alpha$ alone.

**Example 2: Relationship Between Weighted Sum of Errors ($\Psi$) and $\alpha$, Frequentist Analysis.**

Fixing $n_1 = 64$, $\delta_0 = 4$ and $\sigma = 8$ and further, for illustration, we assume that $\omega = 3$, that is we assume the cost of a type I error is three times that of a type II error we can explore the relationship between $\Psi$ and $\alpha$ represented by (7). The red curve in Figure 1 illustrates their relationship. Clearly a minimum, in terms of $\alpha$, exists and this together with its associated β, are the optimum values.

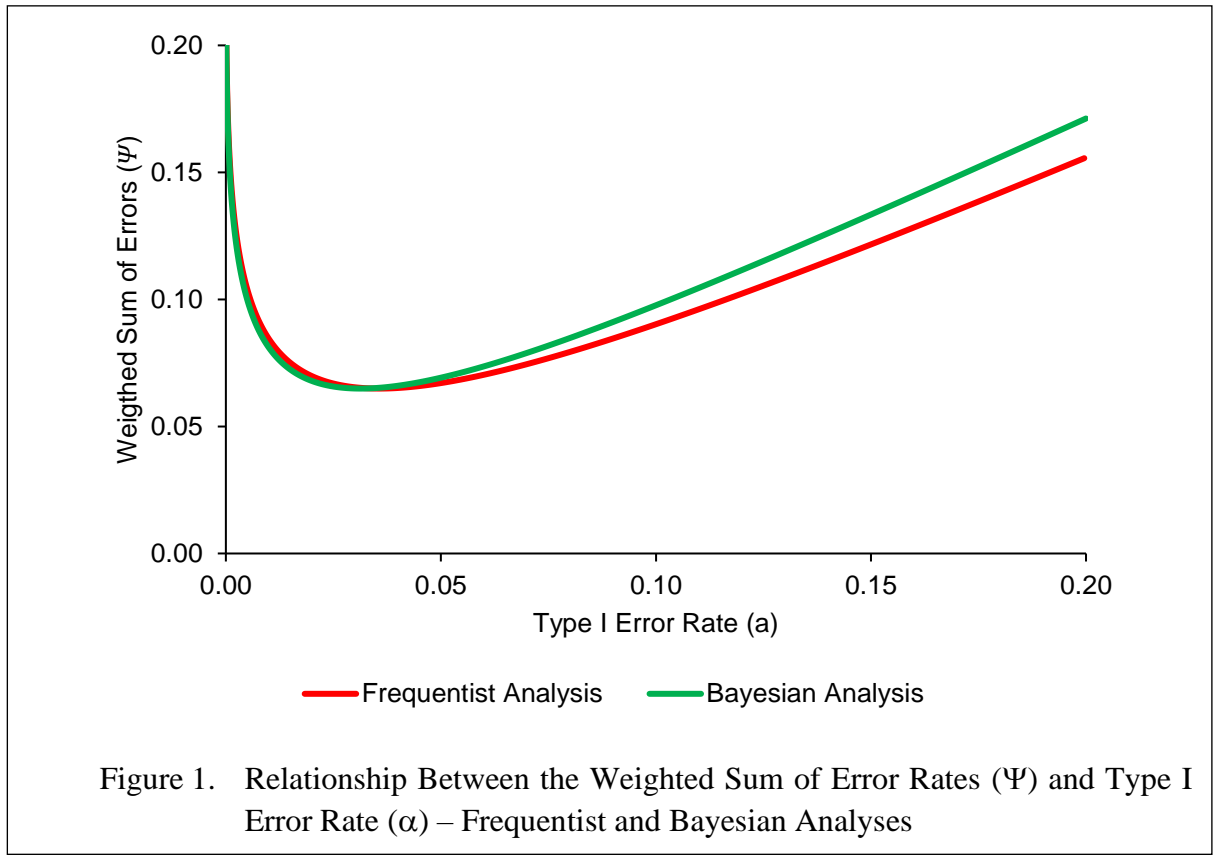

Figure 1. Relationship Between the Weighted Sum of Error Rates ($\Psi$) and Type I Error Rate ($\alpha$) – Frequentist and Bayesian Analyses

To determine the optimal values of α and β, Mudge et al (2012) use numerical search while Maier and Lakens (2022) use numerical optimisation. In contrast, Grieve (2015) provides an analytic solution based on the normal theory assumptions given by Theorem 2.1 (see also the online appendix to Isakov et al, 2019).

**Theorem 2.1.**

The values of $\alpha$ and $\beta$ minimising

$$\Psi = \frac{\omega\alpha + \beta}{\omega + 1}$$

are

$$\alpha = \Phi\left(-\frac{\ln(\omega)}{\theta} - \frac{\theta}{2}\right), \quad \beta = 1 - \Phi\left(-\frac{\ln(\omega)}{\theta} + \frac{\theta}{2}\right)$$

with minimum value

$$\Psi = \frac{\omega\Phi\left(-\frac{\ln(\omega)}{\theta} - \frac{\theta}{2}\right) + \Phi\left(\frac{\ln(\omega)}{\theta} - \frac{\theta}{2}\right)}{\omega + 1}. \tag{8}$$

**Proof.** See Supplementary Material.

If $\omega = 1$, then $\alpha = \beta = \Phi(-\theta/2)$ which defines Arrow's (1960) equal-probability test.

**Example 3: Optimal Values for $\alpha$ and $\beta$, Frequentist Analysis.**

Using the assumptions outlined above for the RLS study design, the value of $\theta$ is 2.8284 which gives the following optimal values for α and β,

$$\alpha = \Phi\left(-\frac{\ln(3)}{2.8284} - \frac{2.8284}{2}\right) = 0.0357 \quad \text{and} \quad \beta = 1 - \Phi\left(-\frac{\ln(3)}{2.8284} + \frac{2.8284}{2}\right) = 0.1525$$

so that the minimum weighted sum of errors is

$$\Psi = \frac{3 \times 0.0357 + 0.1525}{4} = 0.0649.$$

Whilst we consider only the normal case with known variance in this paper, as Grieve (2015) has shown a more general result is available. Concretely if the power of a given test $T$, with associated critical value $t_\alpha$ only depends on a non-central distribution function $G(\theta, t)$ characterised by a single non-centrality parameter $\theta$, then we can write $\Psi$ in the form,

$$\Psi = \frac{\omega G(0, t_\alpha) + 1 - G(\theta, t_\alpha)}{\omega + 1}$$

from which the minimum $\alpha$ is the solution to

$$\omega = \frac{g(\theta, t_\alpha)}{g(0, t_\alpha)} \tag{9}$$

where $g(\theta, t)$ is the non-central density associated with $T$. In Grieve (2015) this result is used to calculate the relative weight for a t-density to achieve a given type I error rate, for example 0.05.

What Theorem 2.1 shows is how for fixed $n_1, \delta_0$ and $\omega$ we can determine the values of $\alpha$ and $\beta$ which minimise the cost-weighted average type I and type II error rates. In planning a clinical trial, we may be more interested in determining the appropriate sample size indeed Mudge et al (2012a) suggest that this could be achieved by requiring that the cost-weighted sum of errors is minimised. Grieve (2015) has shown how to do this for a single-arm clinical study which can be extended to two-arm studies considered here. The optimal weighted sum given by (8) is a function of both $\omega$ and the non-centrality parameter $\theta$, defined in (6). If we were to require that $\Psi$ not exceed a pre-determined value $\Psi_0$, then for a fixed $\omega$ we can determine the optimal $\theta^*$ using the method of *regula falsi* and solve for $n_1$ giving,

$$n_1 = \frac{2\sigma^2 (\theta^*)^2}{\delta_0^2}$$

which as Grieve (2015) pointed out has the same form as (4).

**Example 4: Determining the Sample Size to Control Weighted Sum of Errors**

Suppose in the RLS study we want to control the weighted sum of errors, $\Psi$, to be no more than 0.05 when the type one error is three times more important than a type two error ($\omega = 3$). Figure 2 shows the relationship between $\Psi$ and $\theta^2$ and using the method of *regula falsi* we can find that a value of $\theta^2 = 9.6487$ corresponds to $\Psi = 0.05$. The previous assumptions $\delta_0 = 4$ and $\sigma = 8$ therefore imply that the required sample size per arm is given by,

$$n_1 = \frac{2 \times 64 \times 9.6487}{16} = 77.2$$

which we round up to 78 per arm and for which the optimal values of $\alpha$ and $\beta$ are 0.0279 and 0.1133 respectively.

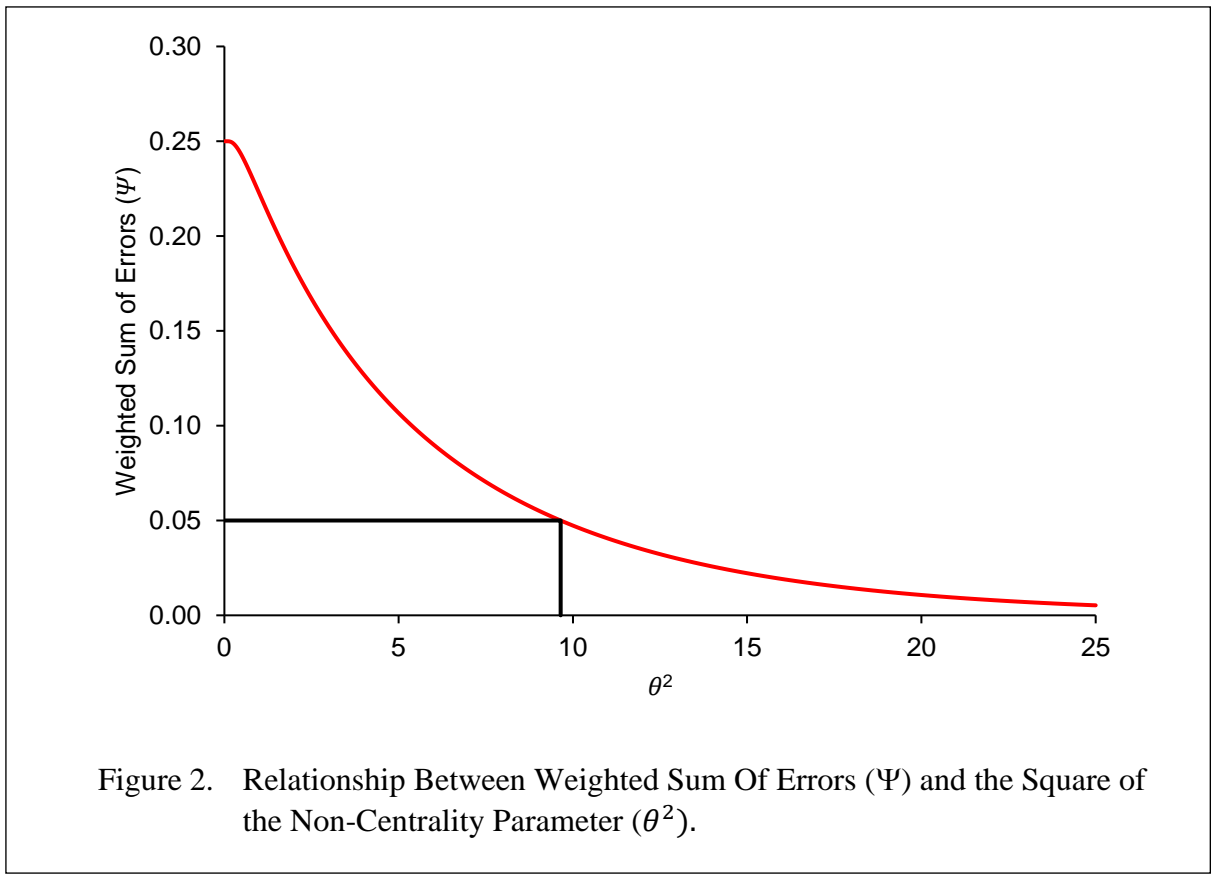

Figure 2. Relationship Between Weighted Sum Of Errors ($\Psi$) and the Square of the Non-Centrality Parameter ($\theta^2$).

**2.2 Bayesian Analysis.**

For a Bayesian, the full probability model has two components, the likelihood of the data given the parameters of the model, and the prior distribution for the parameters.

**2.2.1 Informative Prior for the True Treatment Difference - δ**

For our purposes the likelihood of the simple two arm clinical trial described in Section 2.1 is

$$\hat{\delta} \sim N\left(\delta, \frac{2\sigma^2}{n_1}\right). \tag{10}$$

Throughout we will assume that our expected knowledge, or belief, about the treatment effect before the trial is represented by $\delta_0$, but acknowledge uncertainty about the treatment effect. The uncertainty we will express by a normal distribution for the treatment effect with prior mean $\delta_0$ and variance equivalent to $n_0$ patients. In other words, our prior belief, or prior uncertainty, about the treatment effect can be represented by the probability density function,

$$p(\delta) = N\left(\delta_0, \frac{2\sigma^2}{n_0}\right). \tag{11}$$

A Bayesian analysis follows by combining (10) and (11) by Bayes' Theorem to provide the posterior distribution,

$$p(\delta|\hat{\delta}) \sim N\left(\frac{n_1\hat{\delta} + n_0\delta_0}{n_1 + n_0}, \frac{2\sigma^2}{n_1 + n_0}\right).$$

The standard approach in drug development is to pre-define a decision criterion indicating success. Here, the Bayesian criterion which defines a successful trial is that the lower $1 - \alpha$ credible interval (LCI) for the treatment effect,

$$\frac{n_1\hat{\delta} + n_0\delta_0}{n_1 + n_0} - Z_{1-\alpha}\sqrt{\frac{2\sigma^2}{n_1 + n_0}}$$

excludes 0, or equivalently that,

$$\hat{\delta} > -\frac{n_1 + n_0}{n_1} Z_\alpha \sqrt{\frac{2\sigma^2}{n_1 + n_0}} - \frac{n_0}{n_1}\delta_0. \tag{12}$$

**Lemma 2.1**

For a clinical trial with the decision rule given by (12) under the null hypothesis $H_0$, $p(\hat{\delta}) \sim N(0, 2\sigma^2/n_1)$ the probability of a type I error (T1E) is

$$P(T1E) = \Phi\left(\sqrt{\frac{1}{1-f_0}}Z_\alpha + \sqrt{\frac{f_0}{1-f_0}}Z_0\right) \quad (13)$$

where $f_0 = \dfrac{n_0}{n_0 + n_1}$ and $Z_0 = \sqrt{\dfrac{n_0}{2}}\dfrac{\delta_0}{\sigma}$.

**Proof.** See Supplementary Material.

If we were interested in controlling the type I absolutely at a given level, for example at a one-sided $\varepsilon/2$, then from (13) we would need,

$$Z_{1-a} = \sqrt{1-f_0}Z_{1-\varepsilon/2} + \sqrt{f_0}Z_0$$

which implies that the decision criterion (25) would become,

$$\hat{\delta} > \frac{\sqrt{2}\sigma}{\sqrt{n_1}}Z_{1-\varepsilon/2}$$

and as Grieve (2016) has pointed out this would require us to totally discount all the prior information since this latter decision criterion is the standard frequentist one.

**Lemma 2.2**

For a clinical trial with the decision rule given by (12) under the alternative hypothesis $H_1$, $p(\hat{\delta}) \sim N(\delta_0, 2\sigma^2/n_1)$ the probability of a type II error (T2E) is

$$P(T2E) = \Phi\left(\sqrt{\frac{1}{1-f_0}}Z_\alpha + \sqrt{\frac{f_0}{1-f_0}}Z_0 + \theta\right) \quad (14)$$

where $\theta$ is given by (6).

**Proof.** See Supplementary Material.

The result in (14) is directly related to the concept of conditional Bayesian power (CBP) in Grieve (2022) Section 5.2, since $1 - P(T2E) = CBP$ (defined in (5.3) in that reference).

From (13) and (14) the weighted sum of errors is

$$\Psi = \frac{\omega\Phi\left(\sqrt{\frac{1}{1-f_0}}Z_\alpha + \sqrt{\frac{f_0}{1-f_0}}Z_0\right) + 1 - \Phi\left(\sqrt{\frac{1}{1-f_0}}Z_\alpha + \sqrt{\frac{f_0}{1-f_0}}Z_0 + \theta\right)}{\omega + 1}. \quad (15)$$

**Example 2 (Continued): Relationship Between the Weighted Sum of Errors ($\Psi$) and the Decision Criterion Parameter $\alpha$, Bayesian Analysis.**

In their Bayesian approach to the design of the RLS trial, Muirhead and Soaita (2013) assume the prior distribution for $\delta$ is $N(4,64)$, implying the prior mean is $\delta_0 = 4$ IRLS units and the variance corresponds to a per arm sample size of $n_0 = 2$ patients, so that $Z_0 = 0.5$. If $\omega = 3$, we can explore the relationship between $\Psi$ and $\alpha$ (15) and compare it to the relationship studied in Section 2.1. The green curve in Figure 1 illustrates this relationship. Again, a minimum decision criterion ($\alpha$) exists with its value being given by the following theorem.

**Theorem 2.2.**

The value of $\alpha$ minimising $\Psi$ in (15) is,

$$\alpha = \Phi\left(\sqrt{1-f_0}\left(-\frac{ln(\omega)}{\theta} - \frac{\theta}{2}\right) - \sqrt{f_0}Z_0\right).$$

**Proof.** See Supplementary Material.

**Example 3 (Continued): Optimal Value for $\alpha$, Bayesian Analysis.**

Returning to the assumptions of the RLS trial design we have:

$$f_0 = 0.0303, \quad Z_0 = 0.5, \quad \theta = 2.8284$$

so that

$$\alpha = \Phi\left(0.98743\left(-\frac{ln(3)}{2.8284} - \frac{2.8284}{2}\right) - 0.17408 \times 0.5\right) = 0.0313.$$

In this case the optimum value is almost the same as the frequentist value of 0.0357. This is not unexpected and arises because $f_0$ is small and $\sqrt{1-f_0}$ is close to 1, a consequence of the limited prior information as measured by $n_0$ being used.

In contrast, were the amount of prior information as measured by $n_0$ much larger, the value of $\sqrt{1-f_0}$ will no longer be close to 1, so that the optimal values in the two cases will now be much further apart. However, the result in (16) implies that the minimum value of $\Psi$ in the Bayesian case is identical to the frequentist value a result shown by Walley and Grieve (2021) for a single arm trial but is more generally true. This result is illustrated graphically in Figure 1 where the weighted sum of errors at their respective optimal values are the same.

**2.2.2 Informative Prior for the True Placebo Effect - $\pi$**

Whilst the use of Bayesian approaches in drug development is less controversial than it once was, there remain concerns about their use in phase III clinical trials. One aspect of Bayesian approaches which has gained traction is the use of historical control data to augment control groups. The use of historical data in this way is certainly not new, having been proposed for use in pre-clinical toxicology studies (Dempster et al, 1983) and in clinical contexts (Pocock, 1976). More recently Spiegelhalter et al (2003) and Neuenschwander et al (2010) renewed interest in using historical control data and have reiterated the advantages of this approach.

The impulse for the recent developments in the use of historical data, particularly from control arms, has come from pharmaceutical statisticians. For example, in addition to the predictive priors proposed by Neuenschwander et al (2010), Ibrahim and Chen (2000) introduced power priors with issues of its original formulation being addressed by Duan and Smith (2006), Duan et al (2006) and Neuenschwander et al (2009) and Hobbs et al (2012) proposed commensurate priors. Recent reviews have been provided by Viele et al (2014) and Lim et al (2018). This work is exclusively Bayesian but there are frequentist analogues. Tarone (1982), for example, constructed a test for trend in proportions based around a Cochran-Armitage statistic in which an adjustment to the concurrent control data is made based on the available historical data.

The use of historical data in confirmatory trials is not the focus of the current paper, it is nonetheless useful to look at balancing type I and type II errors in such cases. Here, we assume that based on historical data an informative prior is available for the true placebo response mean, or standard of care if appropriate, but a vague prior is used for the treatment difference, or effect. Concretely, suppose that the prior for the placebo mean is $N(\pi_0, \sigma^2/n_0)$, and that an improper, flat prior is used for the treatment effect, $\delta$. The likelihood for the data is,

$$p\begin{pmatrix}\bar{x}_P\\\bar{x}_A\end{pmatrix} \sim N\left(\begin{pmatrix}\pi\\\pi+\delta\end{pmatrix}, \sigma^2\begin{pmatrix}1/n_P & 0\\0 & 1/n_A\end{pmatrix}\right) \quad (16)$$

then we can use standard results from Lindley and Smith (1972) to show that the posterior distribution for $\delta$ is

$$p(\delta|\bar{x}_P, \bar{x}_A) \sim N\left(\bar{x}_A - \frac{n_0\pi_0 + n_P\bar{x}_p}{n_0 + n_P}, \sigma^2\left(\frac{1}{n_A} + \frac{1}{n_0 + n_P}\right)\right)$$

.

The LCI for the treatment effect, $\delta$, is,

$$\bar{x}_A - \frac{n_0\pi_0 + n_P\bar{x}_p}{n_0 + n_P} - z_{1-\alpha}\sigma\sqrt{\left(\frac{1}{n_A} + \frac{1}{n_0 + n_P}\right)}.$$

so that the condition LCI > 0 implies that

$$\bar{x}_A - \frac{n_P\bar{x}_p}{n_0 + n_P} > \frac{n_0\pi_0}{n_0 + n_P} + z_{1-\alpha}\sigma\sqrt{\left(\frac{1}{n_A} + \frac{1}{n_0 + n_P}\right)}. \quad (17)$$

The distribution of $\bar{x}_A - n_P\bar{x}_p/(n_0 + n_P)$ under the null and alternative hypotheses is required so that the probabilities of both type I and type II errors can be determined. From (16),

$$\bar{x}_A - \frac{n_P\bar{x}_p}{n_0 + n_P} \sim N\left(\frac{n_0\pi}{n_0 + n_P} + \delta, \sigma^2\left(\frac{1}{n_A} + \frac{n_P}{(n_0 + n_P)^2}\right)\right) \quad (18)$$

The probability of a type I error (T1E) is then given by the following lemma.

**Lemma 2.3.**

Based on the decision criterion (22) the *P(T1E)* is given by the formula,

$$P(T1E) = \Phi\left(\frac{\frac{n_0(\pi_0 - \pi)}{n_0 + n_P} + z_{1-\alpha}\sigma\sqrt{\frac{1}{n_A} + \frac{1}{n_0 + n_P}}}{\sigma\sqrt{\frac{1}{n_A} + \frac{n_P}{(n_0 + n_P)^2}}}\right) \quad (19)$$

**Proof.** See Supplementary Material.

Similarly, the probability of a type II error (T2E) is then given by the following lemma.

**Lemma 2.4.**

Based on the decision criterion (17) the *P(T2E)* is given by the formula,

$$P(T2E) = 1 - \Phi\left(\frac{\frac{n_0(\pi_0 - \pi)}{n_0 + n_P} + z_{1-\alpha}\sigma\sqrt{\frac{1}{n_A} + \frac{1}{n_0 + n_P}}}{\sigma\sqrt{\frac{1}{n_A} + \frac{n_P}{(n_0 + n_P)^2}}} + \theta\right), \theta = \frac{\delta}{\sigma\sqrt{\frac{1}{n_A} + \frac{n_P}{(n_0 + n_P)^2}}} \quad (20)$$

**Proof.** See Supplementary Material.

From (19) and (20) the weighted sum of errors is

$$\Psi =* \frac{\omega\Phi\left(\frac{\frac{n_0(\pi_0 - \pi)}{n_0 + n_P} + z_{1-\alpha}\sigma\sqrt{\frac{1}{n_A} + \frac{1}{n_0 + n_P}}}{\sigma\sqrt{\frac{1}{n_A} + \frac{n_P}{(n_0 + n_P)^2}}}\right) + 1 - \Phi\left(\frac{\frac{n_0(\pi_0 - \pi)}{n_0 + n_P} + z_{1-\alpha}\sigma\sqrt{\frac{1}{n_A} + \frac{1}{n_0 + n_P}}}{\sigma\sqrt{\frac{1}{n_A} + \frac{n_P}{(n_0 + n_P)^2}}} + \theta\right)}{\omega + 1} \quad (21)$$

As in the previous section, (21) is structurally identical to (8) so that the value of $\alpha$ which minimises the weighted sum of costs can be determined from setting,

$$\frac{\frac{n_0(\pi_0 - \pi)}{n_0 + n_P} + z_{1-\alpha}\sigma\sqrt{\frac{1}{n_A} + \frac{1}{n_0 + n_P}}}{\sigma\sqrt{\frac{1}{n_A} + \frac{n_P}{(n_0 + n_P)^2}}}$$

equal to $-\frac{\ln(\omega)}{\theta} - \frac{\theta}{2}$ and solving for $\alpha$.

In (23) and (24) the probabilities depend on the "true" value of $\pi$, the population mean for the placebo response. Walley and Grieve (2021) proposed to remove this dependency by taking account of a design prior for the true placebo mean. Suppose that the design prior for $\pi$ is,

$$p(\pi) \sim N\left(\pi_{00}, \frac{\sigma^2}{n_{00}}\right)$$

then the relationship between conditional and marginal normal distributions implies that the unconditional predictive distribution of $\bar{x}_A - \frac{n_P \bar{x}_p}{n_0 + n_P}$ is

$$\bar{x}_A - \frac{n_P \bar{x}_p}{n_0 + n_P} \sim N\left(\frac{n_0 \pi_{00}}{n_0 + n_P} + \delta, \sigma^2\left(\frac{1}{n_A} + \frac{n_P}{(n_0 + n_P)^2} + \frac{n_P^2}{n_{00}(n_0 + n_P)^2}\right)\right)$$

so that

$$P(T1E) = \Phi\left(\frac{\frac{n_0(\pi_0 - \pi_{00})}{n_0 + n_P} + z_{1-\alpha}\sigma\sqrt{\frac{1}{n_A} + \frac{1}{n_0 + n_P}}}{\sigma\sqrt{\frac{1}{n_A} + \frac{n_P}{(n_0 + n_P)^2} + \frac{n_P^2}{n_{00}(n_0 + n_P)^2}}}\right)$$

and

$$P(T2E) = 1 - \Phi\left(\frac{\frac{n_0(\pi_0 - \pi_{00})}{n_0 + n_P} + z_{1-\alpha}\sigma\sqrt{\frac{1}{n_A} + \frac{1}{n_0 + n_P}}}{\sigma\sqrt{\frac{1}{n_A} + \frac{n_P}{(n_0 + n_P)^2} + \frac{n_P^2}{n_{00}(n_0 + n_P)^2}}} + \theta\right)$$

where $\theta = \dfrac{\delta}{\sigma\sqrt{\dfrac{1}{n_A} + \dfrac{n_P}{(n_0 + n_P)^2} + \dfrac{n_P^2}{n_{00}(n_0 + n_P)^2}}}.$

Once again, the structural form of $P(T1E)$ and $P(T2E)$ allows us to utilise the previous results to determine the optimal value for $\alpha$.

## 3. Balancing Type I and Type II Errors When the Hypotheses are Composite.

In developing the approaches in Section 2 we assumed that the null and alternative hypotheses are simple, point hypotheses. Such an assumption is not always appropriate. For example, in bioequivalence testing if $\mu_T$ denotes the population mean area under the concentration-time profile (AUC) for the Test drug and $\mu_R$ denotes the population mean AUC for Reference drug. Then to demonstrate bioequivalence, the null hypotheses to be tested are,

$$H_0: \frac{\mu_T}{\mu_R} \leq \rho_L \quad \text{or} \quad H_0: \frac{\mu_T}{\mu_R} \geq \rho_U$$

versus the alternative hypothesis

$$\rho_L < \frac{\mu_T}{\mu_R} < \rho_U$$

Traditionally, the control of the type I error takes place at the boundary(ies) of the null hypothesis, in the bioequivalence example at $\rho_L$ and $\rho_U$. In this section we take a different approach utilising ideas from Probability of Success (*PoS*), see for example Grieve (2022), also referred to as assurance (O'Hagan et al, 2005) or average power (Spiegelhalter and Freedman, 1988). The argument for taking this approach is made in Pericchi and Pereira (2013).

### 3.1 Frequentist Analysis.

The specific composite null and alternative hypotheses which interest us are $H_0: \delta \leq 0$ and $H_1: \delta > 0$. A test of $H_0$ against $H_1$ is sometimes referred to as a "test for direction" in contrast to a "test for existence". The former is commonly associated with a Bayesian approach in which we are primarily interested in estimating a treatment effect, whilst the latter is more commonly associated with a comparison of simple, point hypotheses using Bayes factors (Marsman and Wagenmakers, 2017). In our development we favour the former.

One way of thinking about *PoS* is to determine the average of the standard power function with respect to the prior distribution of the treatment effect. By rewriting the power function as an integral and changing the order of integration *PoS* can be expressed as the probability that $\hat{\delta}$ exceeds the standard decision criterion calculated with respect to its unconditional, marginal predictive distribution, $p(\hat{\delta})$ giving,

$$PoS = \int_{\frac{\sqrt{2}\sigma Z_{1-\alpha}}{\sqrt{n_1}}}^{\infty} p(\hat{\delta})d\hat{\delta}.$$

Kunzmann et al (2020) point out that the usual definition of *PoS* includes rejections based on values of the treatment effect which are "irrelevant", and in some cases, belong to the region of the null hypothesis. To see this, they provide the following decomposition of the *PoS*:

$$PoS = Pr\left(\hat{\delta} > \frac{Z_{1-\alpha}\sqrt{2}\sigma}{\sqrt{n_1}} \wedge \delta > \delta_{MCID}\right) + Pr\left(\hat{\delta} > \frac{Z_{1-\alpha}\sqrt{2}\sigma}{\sqrt{n_1}} \wedge 0 < \delta < \delta_{MCID}\right)$$
$$+ Pr\left(\hat{\delta} > \frac{Z_{1-\alpha}\sqrt{2}\sigma}{\sqrt{n_1}} \wedge \delta \leq 0\right) = (A) + (B) + (C).$$

.

where $\delta_{MCID}$ is the minimally clinically important difference (MCID). $(C)$ is calculated under the composite null hypothesis which we term the average Type I error rate, $Ave(T1E)$.

**Lemma 3.1**

The $Ave(T1E)$ is given by the formula,

$$Ave(T1E) = B\left(-Z_0, -\sqrt{f_0}(Z_{1-\alpha} - Z_1), -\sqrt{1-f_0}\right). \quad (22)$$

**Proof.** See Supplementary Material.

The expression (22) is identical to (2.19) in Grieve (2022) in which it represents $(C)$, see also Liu (2010) and O'Hagan et al (2005).

**Lemma 3.2**

The $Ave(T2E)$ is given by the formula,

$$Ave(T2E) = B\left(Z_0, \sqrt{f_0}(Z_{1-\alpha} - Z_1), -\sqrt{1-f_0}\right). \quad (23)$$

**Proof.** See Supplementary Material.

This latter result has links to a proposal by Lan and Wittes (2012) to use a truncated (at zero) normal prior distribution for the treatment effect to calculate the *PoS* and to the conditional expected power approach to determining the sample size of a clinical trial introduced by Ciarleglio and Arendt (2017) in which they consider $Pr\left(\hat{\delta} > \frac{Z_{1-\alpha}\sqrt{2}\sigma}{\sqrt{n_1}} \middle| \delta > 0\right)$ rather than $Pr\left(\hat{\delta} > \frac{Z_{1-\alpha}\sqrt{2}\sigma}{\sqrt{n_1}} \wedge \delta > 0\right)$.

**Example 5: Determining the Average Type I and Type II Errors at a fixed value of the Decision Criterion Parameter $\alpha$, Frequentist Analysis.**

Returning to the RLS trial design. As an example, we assume,

$$n_0 = 2, \quad n_1 = 64, \quad \sigma^2 = 64, \quad \delta_0 = 4,$$

from which $Z_1 = 2.8284$, $f_0 = 0.0303$, $\rho = -0.9847$, $Z_{1-\alpha} = 1.95996$, $Z_0 = 0.5$ and $\sqrt{f_0}(Z_{1-\alpha} - Z_1) = -0.15118$

so that from (28) and (29) the average errors are,

$$Ave(T1E) = B(-0.5, \ 0.15118, -0.9847) = 0.000569,$$
$$Ave(T2E) = B(\ 0.5, -0.15118, -0.9847) = 0.131948.$$

Continuing with the assumption that the cost of a type I error is three times that of a type II error, the weighted average error rate is,

$$\Psi = \frac{3 \times 0.000569 + 0.131948}{4} = 0.0333.$$

From (22) and (23) the weighted average of errors can be expressed as

$$\Psi = \frac{\omega \times B\left(-Z_0, -\sqrt{f_0}(Z_{1-\alpha} - Z_1), -\sqrt{1-f_0}\right) + B\left(Z_0, \sqrt{f_0}(Z_{1-\alpha} - Z_1), -\sqrt{1-f_0}\right)}{\omega + 1}. \quad (24)$$

**Example 6: Relationship Between the Weighted Sum of Errors ($\Psi$) and the Decision Criterion Parameter $\alpha$, Frequentist Analysis.**

The previous analysis of this example determined the weighted average of errors at a fixed value of the decision parameter $\alpha = 0.025$. In Figure 3, the red curve shows the relationship between the weighted average of errors ($\Psi$) and the decision criterion parameter $\alpha$ (24). In marked contrast to Figure 1, there is a clear minimum value despite the weighted sum of errors is relatively stable over a wide range of $\alpha$ values.

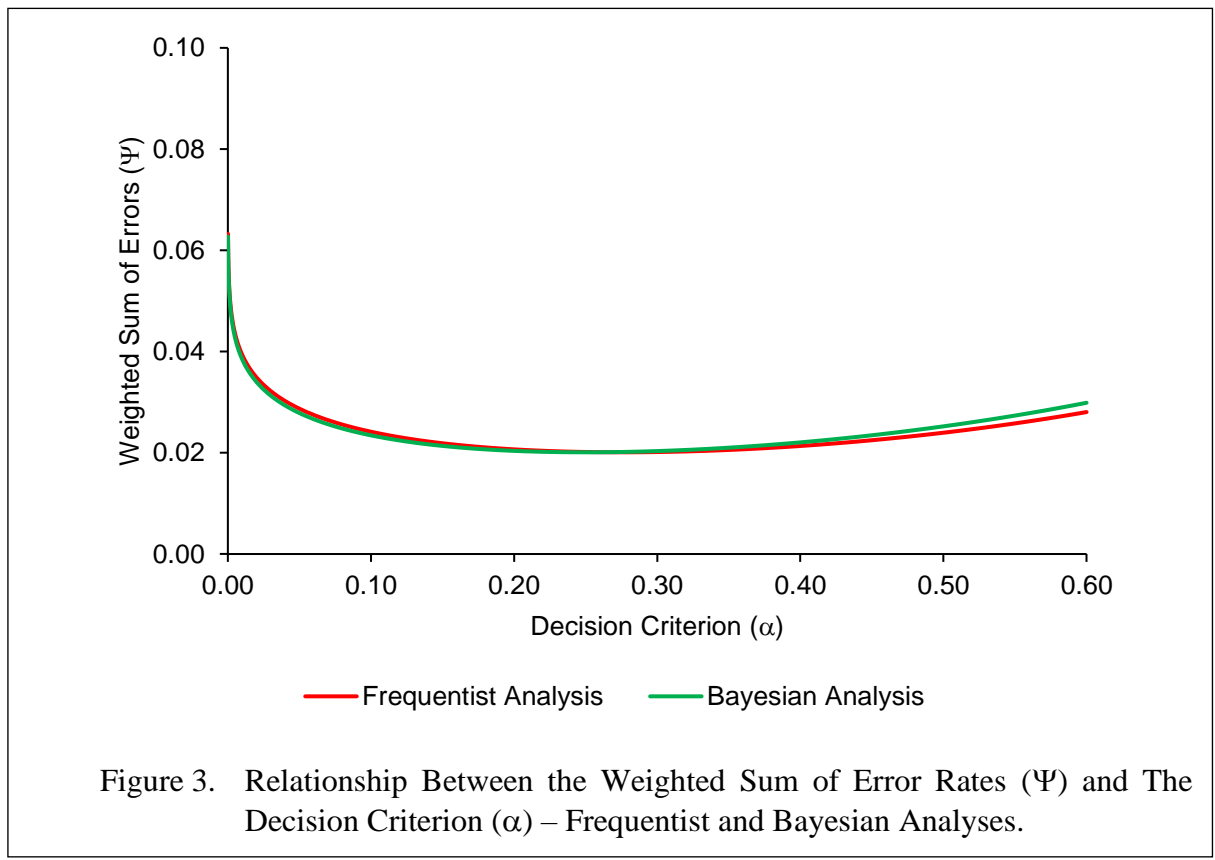

Figure 3. Relationship Between the Weighted Sum of Error Rates ($\Psi$) and The Decision Criterion ($\alpha$) – Frequentist and Bayesian Analyses.

The value of $\alpha$ for which $\Psi$ is a minimum is given in the following theorem.

**Theorem 3.1.**

The value of $\alpha$ minimising $\Psi$ in (24) is.

$$\alpha = 1 - \Phi\left(\frac{\Phi^{-1}\left(\frac{1}{\omega+1}\right)\sqrt{1-\rho^2} + Z_0}{\rho\sqrt{f_0}} + Z_1\right). \tag{25}$$

**Proof.** See Supplementary Material.

**Example 7: Value of $\alpha$ Giving the Minimum Weighted Sum of Errors, Frequentist Analysis**

Returning to our recurring example, we have assumed,

$$n_0 = 2, n_1 = 64, \sigma^2 = 64, \delta_0 = 4, \ \omega = 3$$

from which $Z_1$, $f_0$, $\rho$, $Z_0$ are given in Example 5 so that from (32) the optimum decision criterion is given by

$$\alpha = 1 - \Phi\left(\frac{-0.67449 \times 0.17408 + 0.5}{-0.98473 \times 0.17408} + 2.82843\right) = 0.27540.$$

**3.2 Bayesian Analysis.**

For the composite null and alternative hypotheses $H_0: \delta \leq 0$ and $H_1: \delta > 0$, we use the decision criterion (12) for a standard Bayesian analysis to determine the average type I and type II errors in the same way that we did in Section 3.1.

**Lemma 3.4**

The $Ave(T1E)$ is given by the formula,

$$Ave(T1E) = B\left(-Z_0, -\sqrt{\frac{f_0}{1-f_0}}\left(Z_{1-\alpha} - \sqrt{\frac{1}{1-f_0}}Z_1\right), -\sqrt{1-f_0}\right). \tag{26}$$

**Proof.** See Supplementary Material.

**Lemma 3.5**

The $Ave(T2E)$ is given by the formula,

$$Ave(T2E) = B\left(Z_0, \sqrt{\frac{f_0}{1-f_0}}\left(Z_{1-\alpha} - \sqrt{\frac{1}{1-f_0}}Z_1\right), -\sqrt{1-f_0}\right). \tag{27}$$

**Proof.** See Supplementary Material.

**Example 5 (Continued): Determining the Average Type I and Type II Errors at a fixed value of the Decision Criterion Parameter $\alpha$, Bayesian Analysis.**

We again assume,

$$n_0 = 2, n_1 = 64, \sigma^2 = 64, \delta_0 = 4,$$

from which $Z_1$, $f_0$, $\rho$, $Z_0$ and $Z_{1-\alpha}$ are given in Example 5 and

$$\sqrt{\frac{f_0}{1-f_0}}\left(Z_{1-\alpha} - \sqrt{\frac{1}{1-f_0}}\, Z_1\right) = -0.85243$$

so that from (26) and (27) the average errors for fixed $\alpha$ are,

$$\begin{aligned} Ave(T1E) &= B(-0.5,\ \ 0.16128, -0.9847) = 0.000662, \\ Ave(T2E) &= B(\ \ 0.5, -0.16128, -0.9847) = 0.128062. \end{aligned}$$

Continuing with the assumption that $\omega = 3$ the weighted average error rate is

$$\Psi = \frac{3 \times 0.000662 + 0.128062}{4} = 0.0325.$$

From (26) and (27) the weighted average of errors can then be expressed as

$$\Psi = \frac{\omega \times B\left(-Z_0, -\sqrt{f_0}(Z_{1-\alpha} - Z_1), -\sqrt{1-f_0}\right) + B\left(Z_0, \sqrt{f_0}(Z_{1-\alpha} - Z_1), -\sqrt{1-f_0}\right)}{\omega + 1}. \quad (28)$$

**Example 6 (Continued): Relationship Between the Weighted Sum of Errors ($\Psi$) and the Decision Criterion Parameter $\alpha$, Bayesian Analysis.**

With the Muirhead and Soaita (2013) assumptions and $\omega = 3$ the relationship between $\Psi$ and $\alpha$ (28) is given by the green curve in Figure 3. Again, a minimum exists, however the curves for the frequentist and Bayesian analyses are closer to one another for large values of the decision criterion parameter in this figure than in the corresponding figure, Figure 1.

**Theorem 3.2.**

The value of $\alpha$ minimising $\Psi$ in (28) is,

$$\alpha = 1 - \Phi\left(\frac{\Phi^{-1}\left(\frac{1}{\omega+1}\right)\sqrt{1-\rho^2} + Z_0}{\rho\sqrt{\frac{f_0}{1-f_0}}} + \sqrt{\frac{1}{1-f_0}}\, Z_1\right). \quad (29)$$

**Proof.** See Supplementary Material.

**Example 7 (Continued): Value of $\alpha$ Giving the Minimum Weighted Sum of Errors ($\Psi$), Bayesian Analysis**

Returning to our recurring example, we have assumed,

$$n_0 = 2, n_1 = 64, \sigma^2 = 64, \delta_0 = 4,\ \ \omega = 3$$

from which $Z_1$, $f_0$, ρ and $Z_0$ are given in Example 5 so that from (29) the optimum decision criterion is given by

$$\alpha = 1 - \Phi\left(\frac{-0.67449 \times 0.17408 + 0.5}{-0.98473 \times 0.17678} + 1.01551 \times 2.82843\right) = 0.25.$$

**4. Determining the Ratio of Costs.**

A referee has raised the issue of how ratio of costs (ω) is to be determined.

**5. Discussion.**

In this paper we have provided an overview of approaches to minimising the sum, or weighted sum, of the type I and type II error rates in planning clinical trials. In contrast, the standard approach is to fix the type I error rate and then to minimise the type II error at the alternative. Whilst the ideas underpinning the approaches considered here are over 60 years old, they have until recently not found usage in pharmaceutical drug development, or medical research more generally.

However, changes are afoot. In a series of papers Lo and colleagues have considered how consideration of the prevalence and severity of disease can, and perhaps should, influence the balance between type I and type II errors (Chaudhuri et al, 2018; Isakov et al, 2019; Chaudhuri et al, 2020). In the context of a disease which is invariably fatal, it is not surprising that avoiding a type II error may be more important than avoiding a type I error. This contrasts with say, hypertension, for which many drugs are available and the consequences of making a type II error are likely to be less impactful on patients.

Philosophically, this is no different to Laplace's consideration of the number of votes required in a court to convict a prisoner as reported by Neyman and Pearson (1933). Neyman and Pearson ask whether it is more critical to find an innocent person guilty than to find a guilty person not guilty? They point out that it will depend on the consequences of the error, for example whether a guilty person will be executed or fined.

**Appendix – Proofs of Theorems**

**Theorem 2.1.**

The values of $\alpha$ and $\beta$ minimising

$$\Psi = \frac{\omega\alpha + \beta}{\omega + 1}$$

are

$$\alpha = \Phi\left(-\frac{\ln(\omega)}{\theta} - \frac{\theta}{2}\right), \quad \beta = 1 - \Phi\left(-\frac{\ln(\omega)}{\theta} + \frac{\theta}{2}\right)$$

*with minimum value*

$$\Psi = \frac{\omega\Phi\left(-\frac{\ln(\omega)}{\theta} - \frac{\theta}{2}\right) + \Phi\left(\frac{\ln(\omega)}{\theta} - \frac{\theta}{2}\right)}{\omega + 1}.$$

*Proof.*

*Rewriting (7) as*

$$\Psi = \frac{\omega\Phi(Z_\alpha) + 1 - \Phi(\theta + Z_\alpha)}{\omega + 1}, \quad (A1)$$

then using the Leibniz integral rule, the optimal value of $Z_\alpha$ providing the minimum weighted sum of errors is the solution to

$$\frac{d\Psi}{dZ_\alpha} = 0 = \omega\phi(Z_\alpha) - \phi(\theta + Z_\alpha)$$

which implies

$$\omega = \frac{\phi(\theta + Z_\alpha)}{\phi(Z_\alpha)}. \quad (A2)$$

*The solution to (A2) which provides the optimal $\alpha$ and $\beta$ is*

$$Z_\alpha = -\frac{\ln(\omega)}{\theta} - \frac{\theta}{2} \quad (A3)$$

see also the online appendix to Isakov et al (2019). From (A3) we have

$$\alpha = \Phi\left(-\frac{\ln(\omega)}{\theta} - \frac{\theta}{2}\right), \quad \beta = 1 - \Phi\left(-\frac{\ln(\omega)}{\theta} + \frac{\theta}{2}\right)$$

and from (A1) and (A3) the minimum value of $\Psi$ is

$$\Psi = \frac{\omega\Phi\left(-\frac{\ln(\omega)}{\theta} - \frac{\theta}{2}\right) + \Phi\left(\frac{\ln(\omega)}{\theta} - \frac{\theta}{2}\right)}{\omega + 1}.$$

**Lemma 2.1.**

For a clinical trial with the decision rule given by (11) under the null hypothesis $H_0$, $p(\hat{\delta}) \sim N(0, 2\sigma^2/n_1)$ the probability of a type I error (T1E) is

$$P(T1E) = \Phi\left(\sqrt{\frac{1}{1-f_0}} Z_\alpha + \sqrt{\frac{f_0}{1-f_0}} Z_0\right)$$

where $f_0 = \dfrac{n_0}{n_0 + n_1}$ and $Z_0 = \sqrt{\dfrac{n_0}{2}} \dfrac{\delta_0}{\sigma}$.

**Proof.**

Given the success criterion

$$\hat{\delta} > -\frac{n_1 + n_0}{n_1} Z_\alpha \sqrt{\frac{2\sigma^2}{n_1 + n_0}} - \frac{n_0}{n_1} \delta_0$$

the probability of achieving success under the null hypothesis is

$$P(T1E) = \left(\frac{4\pi\sigma^2}{n_1}\right)^{-\frac{1}{2}} \int_{-\frac{n_1+n_0}{n_1} Z_\alpha \sqrt{\frac{2\sigma^2}{n_1+n_0}} - \frac{n_0}{n_1}\delta_0}^{\infty} exp\left(-\frac{n_1 \hat{\delta}^2}{4\sigma^2}\right) d\hat{\delta}$$

$$= \Phi\left(\sqrt{\frac{n_1 + n_0}{n_1}} Z_\alpha + \sqrt{\frac{n_0}{n_1}} Z_0\right) = \Phi\left(\sqrt{\frac{1}{1-f_0}} Z_\alpha + \sqrt{\frac{f_0}{1-f_0}} Z_0\right)$$

[2] where $f_0 = \dfrac{n_0}{n_0 + n_1}$ and $Z_0 = \sqrt{\dfrac{n_0}{2}} \dfrac{\delta_0}{\sigma}$.

**Lemma 2.2.**

For a clinical trial with the decision rule given by (12) under the alternative hypothesis $H_0$, $p(\hat{\delta}) \sim N(\delta_0, 2\sigma^2/n_1)$ the probability of a type I error (T2E) is

$$P(T2E) = \Phi\left(\sqrt{\frac{1}{1-f_0}} Z_\alpha + \sqrt{\frac{f_0}{1-f_0}} Z_0 + \theta\right)$$

where $\theta$ is given by (6).

*Proof.*

Given the success criterion

$$\hat{\delta} > -\frac{n_1 + n_0}{n_1} Z_\alpha \sqrt{\frac{2\sigma^2}{n_1 + n_0}} - \frac{n_0}{n_1}\delta_0$$

the probability of achieving success under the alternative hypothesis is

$$P(T2E) = \left(\frac{4\pi\sigma^2}{n_1}\right)^{-\frac{1}{2}} \int_{-\frac{n_1+n_0}{n_1}Z_\alpha \sqrt{\frac{2\sigma^2}{n_1+n_0}} - \frac{n_0}{n_1}\delta_0}^{\infty} exp\left(-\frac{n_1(\hat{\delta} - \delta_0)^2}{4\sigma^2}\right) d\hat{\delta}$$

$$= \Phi\left(\sqrt{\frac{n_1 + n_0}{n_1}} Z_\alpha + \sqrt{\frac{n_0}{n_1}} Z_0 + \theta\right) = \Phi\left(\sqrt{\frac{1}{1 - f_0}} Z_\alpha + \sqrt{\frac{f_0}{1 - f_0}} Z_0 + \theta\right)$$

**Theorem 2.2.**

The value of $\alpha$ minimising $\Psi$ in (15) is

$$\alpha = \Phi\left(\sqrt{1 - f_0}\left(-\frac{ln(\omega)}{\theta} - \frac{\theta}{2}\right) - \sqrt{f_0}Z_0\right).$$

*Proof.*

The weighted sum of errors (15) is structurally identical to (7). Therefore, if we set

$$\sqrt{\frac{1}{1 - f_0}} Z_\alpha + \sqrt{\frac{f_0}{1 - f_0}} Z_0 = -\frac{ln(\omega)}{\theta} - \frac{\theta}{2}$$

in which the right-hand side is the optimal solution (A3), we can re-arrange it to give

$$Z_\alpha = \sqrt{1 - f_0}\left(-\frac{ln(\omega)}{\theta} - \frac{\theta}{2}\right) - \sqrt{f_0}Z_0$$

and hence

$$\alpha = \Phi\left(\sqrt{1 - f_0}\left(-\frac{ln(\omega)}{\theta} - \frac{\theta}{2}\right) - \sqrt{f_0}Z_0\right).$$

**Lemma 2.3.**

Based on the decision criterion (17) the *P(T1E)* is given by the formula,

$$P(T1E) = \Phi\left(\frac{\frac{n_0(\pi_0 - \pi)}{n_0 + n_P} + z_{1-\alpha}\sigma\sqrt{\frac{1}{n_A} + \frac{1}{n_0 + n_P}}}{\sigma\sqrt{\frac{1}{n_A} + \frac{n_P}{(n_0 + n_P)^2}}}\right)$$

**Proof.**

Setting $\delta = 0$ in (18) the predictive distribution of probability of $\bar{x}_A - \frac{n_P \bar{x}_p}{n_0 + n_P}$ under the null hypothesis is

$$\bar{x}_A - \frac{n_P \bar{x}_p}{n_0 + n_P} \sim N\left(\frac{n_0 \pi}{n_0 + n_P} + \delta, \sigma^2 \left(\frac{1}{n_A} + \frac{n_P}{(n_0 + n_P)^2}\right)\right)$$

From this distribution the probability of a type I error (T1E) is given by

$$P(T1E) = P\left(\bar{x}_A - \frac{n_P \bar{x}_p}{n_0 + n_P} > \frac{n_0 \pi_0}{n_0 + n_P} + z_{1-\alpha}\sigma \sqrt{\left(\frac{1}{n_A} + \frac{1}{n_0 + n_P}\right)}\right)$$

$$= \Phi\left(\frac{\frac{n_0(\pi_0 - \pi)}{n_0 + n_P} + z_{1-\alpha}\sigma\sqrt{\frac{1}{n_A} + \frac{1}{n_0 + n_P}}}{\sigma\sqrt{\frac{1}{n_A} + \frac{n_P}{(n_0 + n_P)^2}}}\right)$$

**Lemma 2.4.**

Based on the decision criterion (17) the $P(T2E)$ is given by the formula,

$$P(T2E) = 1 - \Phi\left(\frac{\frac{n_0(\pi_0 - \pi)}{n_0 + n_P} + z_{1-\alpha}\sigma\sqrt{\frac{1}{n_A} + \frac{1}{n_0 + n_P}}}{\sigma\sqrt{\frac{1}{n_A} + \frac{n_P}{(n_0 + n_P)^2}}} + \theta\right)$$

where $\quad \theta = \dfrac{\delta}{\sigma\sqrt{\frac{1}{n_A} + \frac{n_P}{(n_0 + n_P)^2}}}.$

**Proof.**

The distribution of $\bar{x}_A - \frac{n_P \bar{x}_p}{n_0 + n_P}$ under the alternative hypothesis is given by equation (18):

$$\bar{x}_A - \frac{n_P \bar{x}_p}{n_0 + n_P} \sim N\left(\frac{n_0 \pi}{n_0 + n_P} + \delta, \sigma^2 \left(\frac{1}{n_A} + \frac{n_P}{(n_0 + n_P)^2}\right)\right).$$

From this distribution the probability of a type I error (T2E) is given by

$$P(T1E) = P\left(\bar{x}_A - \frac{n_P \bar{x}_p}{n_0 + n_P} < \frac{n_0 \pi_0}{n_0 + n_P} + z_{1-\alpha}\sigma\sqrt{\left(\frac{1}{n_A} + \frac{1}{n_0 + n_P}\right)}\right)$$

$$= 1 - \Phi\left(\frac{\frac{n_0(\pi_0 - \pi)}{n_0 + n_P} + z_{1-\alpha}\sigma\sqrt{\frac{1}{n_A} + \frac{1}{n_0 + n_P}}}{\sigma\sqrt{\frac{1}{n_A} + \frac{n_P}{(n_0 + n_P)^2}}} + \theta\right)$$

**Lemma 3.1.**

The $Ave(T1E)$ is given by the formula

$$Ave(T1E) = B\left(-Z_0, -\sqrt{f_0}(Z_{1-\alpha} - Z_1), -\sqrt{1-f_0}\right).$$

**Proof.**
The expression $Ave(T1E)$ can be written in the form:

$$Ave(T1E) = Pr\left(\hat{\delta} > \frac{Z_{1-\alpha}\sqrt{2}\sigma}{\sqrt{n_1}} \wedge \delta < 0\right) = \int_{-\infty}^{0}\int_{\frac{Z_{1-\alpha}\sqrt{2}\sigma}{\sqrt{n_1}}}^{\infty} p(\hat{\delta}, \delta)\, d\hat{\delta}d\delta.$$

The joint distribution $p(\hat{\delta}, \delta)$ is determined by noting that the distribution of the treatment estimate, $\hat{\delta}$, conditional on the population treatment effect $\delta$ is given by (10), whilst the prior distribution of the population treatment effect is given by (11), so that properties of the bivariate normal distribution imply

$$p(\hat{\delta}, \delta) \sim N\left(\begin{pmatrix}\delta_0\\\delta_0\end{pmatrix}, \frac{2\sigma^2}{n_0}\begin{pmatrix}\frac{(n_0+n_1)}{n_1} & 1\\ 1 & 1\end{pmatrix}\right). \tag{A4}$$

If we transform the variables $\hat{\delta}$ and $\delta$ to $y$ and $x$ respectively utilising the expressions:

$$y = \frac{(\hat{\delta} - \delta_0)\sqrt{n_0 n_1}}{\sqrt{2\sigma^2(n_0+n_1)}}, \quad x = \frac{(\delta - \delta_0)\sqrt{n_0}}{\sqrt{2\sigma^2}} \tag{A5}$$

then $Ave(T1E)$ can be written as

$$\int_{-\infty}^{-Z_0}\int_{\sqrt{f_0}(Z_{1-\alpha}-Z_1)}^{\infty} g(x, y, \rho)\, dy\, dx$$

where $g(x, y, \rho) = \frac{1}{2\pi\sqrt{1-\rho^2}}\exp\left(-\frac{x^2 - 2\rho xy + y^2}{2(1-\rho^2)}\right)$, $Z_1 = \sqrt{\frac{n_1}{2}}\frac{\delta_0}{\sigma}$, $f_0 = \frac{n_0}{n_1+n_0}$ and $\rho = \sqrt{1-f_0}$.

We can use standard properties of the bivariate normal distribution to evaluate $Ave(T1E)$ using existing functions in statistical packages, for example PROBBNRM in SAS, which evaluates

$$B(h, k, \rho) = \int_{-\infty}^{h}\int_{-\infty}^{k} g(x, y, \rho)\, dy\, dx. \tag{A6}$$

In this way,

$$Ave(T1E) = B\left(-Z_0, -\sqrt{f_0}(Z_{1-\alpha} - Z_1), -\sqrt{1-f_0}\right).$$

**Lemma 3.2.**

The $Ave(T2E)$ is given by the formula

$$Ave(T2E) = B\big(Z_0, \sqrt{f_0}(Z_{1-\alpha} - Z_1), -\sqrt{1-f_0}\big).$$

**Proof.**
The expression $Ave(T2E)$ can be written in the form:

$$Ave(T2E) = Pr\left(\hat{\delta} < \frac{Z_{1-\alpha}\sqrt{2}\sigma}{\sqrt{n_1}} \wedge \delta > 0\right) = \int_0^\infty \int_{-\infty}^{\frac{Z_{1-\alpha}\sqrt{2}\sigma}{\sqrt{n_1}}} p(\hat{\delta}, \delta)\, d\hat{\delta}d\delta.$$

Using (A4) and (A5) again, $Ave(T2E)$ can be written as

$$\int_{-Z_0}^\infty \int_{-\infty}^{\sqrt{f_0}(Z_{1-\alpha}-Z_1)} g(x,y,\rho)\,dydx,$$

which in turn can be expressed as

$$Ave(T2E) = B\big(-Z_0, -\sqrt{f_0}(Z_{1-\alpha} - Z_1), -\sqrt{1-f_0}\big).$$

**Lemma 3.3.**

The partial derivatives of the standard bivariate normal cumulative distribution function (A6) are given by the formulae

$$\frac{\partial B(h,k,\rho)}{\partial h} = \phi(h)\Phi\left(\frac{k-\rho h}{\sqrt{1-\rho^2}}\right), \qquad \frac{\partial B(h,k,\rho)}{\partial k} = \phi(k)\Phi\left(\frac{h-\rho k}{\sqrt{1-\rho^2}}\right). \qquad \text{g}$$

*Proof.*

*Using Leibniz' integral rule*

$$\frac{\partial B(h,k,\rho)}{\partial h} = \frac{1}{2\pi\sqrt{1-\rho^2}} \int_{-\infty}^k \exp\left\{-\frac{1}{2(1-\rho^2)}(h^2 - 2hy\rho + y^2)\right\} dy$$

$$= \frac{1}{2\pi\sqrt{1-\rho^2}} \int_{-\infty}^k \exp\left\{-\frac{h^2}{2(1-\rho^2)}\right\} \exp\left\{-\frac{(y-h\rho)^2 - \rho^2 h^2}{2(1-\rho^2)}\right\} dy$$

$$= \frac{\exp\left\{-\frac{h^2}{2}\right\}}{\sqrt{2\pi}\sqrt{2\pi(1-\rho^2)}} \int_{-\infty}^k \exp\left\{-\frac{(y-h\rho)^2}{2(1-\rho^2)}\right\} dy$$

$$= \phi(h)\Phi\left(\frac{k-h\rho}{\sqrt{1-\rho^2}}\right).$$

A similar approach shows that

$$\frac{\partial B(h,k,\rho)}{\partial k} = \phi(k)\Phi\left(\frac{h-k\rho}{\sqrt{1-\rho^2}}\right).$$

**Theorem 3.1.**

The value of $\alpha$ minimising $\Psi$ in (24) is

$$\alpha = 1 - \Phi\left(\frac{\Phi^{-1}\left(\frac{1}{\omega+1}\right)\sqrt{1-\rho^2}+Z_0}{\rho\sqrt{f_0}} + Z_1\right).$$

*Proof.*

Using Lemma 4.3 we determine the partial derivative of $\Psi$ in (24) with respect to $Z_{1-\alpha}$ and set it equal to zero, so that

$$\frac{d\Psi}{dZ_{1-\varepsilon}} = 0 \Rightarrow \sqrt{f_0}\frac{\omega}{\omega+1}\phi\left(-\sqrt{f_0}(Z_{1-\alpha}-Z_1)\right)\Phi\left(\frac{-Z_0+\rho\sqrt{f_0}(Z_{1-\alpha}-Z_1)}{\sqrt{1-\rho^2}}\right)$$
$$= \sqrt{f_0}\frac{1}{\omega+1}\phi\left(\sqrt{f_0}(Z_{1-\alpha}-Z_1)\right)\Phi\left(\frac{Z_0-\rho\sqrt{f_0}(Z_{1-\alpha}-Z_1)}{\sqrt{1-\rho^2}}\right).$$

From which

$$\frac{1}{\omega+1} = \Phi\left(\frac{-Z_0+\rho\sqrt{f_0}(Z_{1-\alpha}-Z_1)}{\sqrt{1-\rho^2}}\right)$$

and this expression in turn can be rearranged to give

$$\alpha = 1 - \Phi\left(\frac{\Phi^{-1}\left(\frac{1}{\omega+1}\right)\sqrt{1-\rho^2}+Z_0}{\rho\sqrt{f_0}} + Z_1\right).$$

**Lemma 3.4.**

The $Ave(T1E)$ is given by the formula

$$Ave(T1E) = B\left(-Z_0, -\sqrt{\frac{f_0}{1-f_0}}\left(Z_{1-\alpha} - \sqrt{\frac{1}{1-f_0}}Z_1\right), -\sqrt{1-f_0}\right).$$

**Proof.**

The $Ave(T1E)$ can be expressed as

$$Ave(T1E) = Pr\left(\hat{\delta} > \frac{Z_{1-\alpha}\sqrt{2}\sigma\sqrt{n_1+n_0}}{n_1} - \frac{n_0}{n_1}\delta_0 \wedge \delta < 0\right)$$

Proceeding as we did in Section 3.2, we can express $Ave(T1E)$ as:

$$Ave(T1E) = \int_{-\infty}^{0} \int_{\frac{Z_{1-\alpha}\sqrt{2}\sigma\sqrt{n_1+n_0}}{n_1}-\frac{n_0}{n_1}\delta_0}^{\infty} p(\hat{\delta},\delta)\,d\delta d\delta$$

where $p(\hat{\delta},\delta)$ is as previously defined. Using the transformations in (A5) $Ave(T1E)$ can be written as

$$Ave(T1E) = \int_{-\infty}^{-Z_0} \int_{\sqrt{\frac{f_0}{1-f_0}}\left(Z_{1-\alpha}-\sqrt{\frac{1}{1-f_0}}Z_1\right)}^{\infty} g(x,y,\rho)\,dy dx$$

so that

$$Ave(T1E) = B\left(-Z_0, -\sqrt{\frac{f_0}{1-f_0}}\left(Z_{1-\alpha} - \sqrt{\frac{1}{1-f_0}}Z_1\right), -\sqrt{1-f_0}\right).$$

**Lemma 3.5.**

The $Ave(T2E)$ is given by the formula

$$Ave(T2E) = B\left(Z_0, \sqrt{\frac{f_0}{1-f_0}}\left(Z_{1-\alpha} - \sqrt{\frac{1}{1-f_0}}Z_1\right), -\sqrt{1-f_0}\right).$$

**Proof.**

The expression $Ave(T2E)$ can be written in the form:

$$Ave(T2E) = Pr\left(\hat{\delta} < \frac{Z_{1-\alpha}\sqrt{2}\sigma\sqrt{n_1+n_0}}{n_1} - \frac{n_0}{n_1}\delta_0 \wedge \delta > 0\right)$$

Proceeding as before, we can express $Ave(T2E)$ as:

$$Ave(T2E) = \int_{0}^{\infty} \int_{-\infty}^{\frac{Z_{1-\alpha}\sqrt{2}\sigma\sqrt{n_1+n_0}}{n_1}-\frac{n_0}{n_1}\delta_0} p(\hat{\delta},\delta)\,d\delta d\delta$$

where $p(\hat{\delta},\delta)$ is as previously defined. Using the transformations in (A5) $Ave(T2E)$ can be written as

$$Ave(T2E) = \int_{-Z_0}^{\infty} \int_{-\infty}^{\sqrt{\frac{f_0}{1-f_0}}\left(Z_{1-\alpha} - \sqrt{\frac{1}{1-f_0}}Z_1\right)} g(x,y,\rho) dy dx$$

so that

$$Ave(T2E) = B\left(Z_0, \sqrt{\frac{f_0}{1-f_0}}\left(Z_{1-\alpha} - \sqrt{\frac{1}{1-f_0}}Z_1\right), -\sqrt{1-f_0}\right).$$

**Theorem 3.2.**

The value of $\alpha$ minimising $\Psi$ in (28) is

$$\alpha = 1 - \Phi\left(\frac{\Phi^{-1}\left(\frac{1}{\omega+1}\right)\sqrt{1-\rho^2} + Z_0}{\rho\sqrt{\frac{f_0}{1-f_0}}} + \sqrt{\frac{1}{1-f_0}}Z_1\right).$$

**Proof.**

Using Lemma 4.3 we determine the partial derivative of $\Psi$ in (28) with respect to $Z_{1-\alpha}$ and set it equal to zero, so that

$$\frac{d\Psi}{dZ_{1-\varepsilon}} = 0 \Rightarrow$$

$$\sqrt{\frac{f_0}{1-f_0}}\frac{\omega}{\omega+1}\phi\left(-\sqrt{\frac{f_0}{1-f_0}}\left(Z_{1-\alpha} - \sqrt{\frac{1}{1-f_0}}Z_1\right)\right)\Phi\left(\frac{-Z_0 + \rho\sqrt{\frac{f_0}{1-f_0}}\left(Z_{1-\alpha} - \sqrt{\frac{1}{1-f_0}}Z_1\right)}{\sqrt{1-\rho^2}}\right)$$

$$= \sqrt{\frac{f_0}{1-f_0}}\frac{1}{\omega+1}\phi\left(-\sqrt{\frac{f_0}{1-f_0}}\left(Z_{1-\varepsilon} - \sqrt{\frac{1}{1-f_0}}Z_1\right)\right)\Phi\left(\frac{Z_0 - \rho\sqrt{\frac{f_0}{1-f_0}}\left(Z_{1-\varepsilon} - \sqrt{\frac{1}{1-f_0}}Z_1\right)}{\sqrt{1-\rho^2}}\right)$$

From which

$$\frac{1}{\omega+1} = \Phi\left(\frac{Z_0 - \rho\sqrt{\frac{f_0}{1-f_0}}\left(Z_{1-\alpha} - \sqrt{\frac{1}{1-f_0}}Z_1\right)}{\sqrt{1-\rho^2}}\right)$$

and this expression in turn can be rearranged to give

$$\alpha = 1 - \Phi\left(\frac{\Phi^{-1}\left(\frac{1}{\omega+1}\right)\sqrt{1-\rho^2} + Z_0}{\rho\sqrt{\frac{f_0}{1-f_0}}} + \sqrt{\frac{1}{1-f_0}}Z_1\right)..$$